\documentclass[nofootinbib,superscriptaddress,showpacs,amssymb,10pt,aps,prd,longbibliography,reprint]{revtex4-1}

\usepackage{graphicx,epsfig,amssymb} 
\usepackage{amsmath,amsfonts, times}
\usepackage{bm} 

\usepackage{epstopdf}
\usepackage[linktocpage,colorlinks]{hyperref}
\usepackage[caption=false]{subfig}
\usepackage[usenames]{color}     
\usepackage{natbib}
\usepackage{soul}

\definecolor{coolblack}{rgb}{0.0, 0.18, 0.39}
\definecolor{darkred}{rgb}{0.5,0,0}
\definecolor{darkgreen}{rgb}{0,0.5,0}
\definecolor{darkblue}{rgb}{0,0,0.5}
\definecolor{lapislazuli}{rgb}{0.15, 0.38, 0.61}
\definecolor{venetianred}{rgb}{0.78, 0.03, 0.08}
\definecolor{bleudefrance}{rgb}{0.19, 0.55, 0.91}
\definecolor{dogwoodrose}{rgb}{0.84, 0.09, 0.41}
\hypersetup{colorlinks=true, citecolor=darkgreen, linkcolor=darkblue, 
urlcolor = blue}

\def\be{\begin{equation}}
\def\ee{\end{equation}}

\newcommand{\bea}{\begin{eqnarray}}
\newcommand{\eea}{\end{eqnarray}}
\newcommand{\ben}{\begin{enumerate}}
\newcommand{\een}{\end{enumerate}}
\newcommand{\bi}{\begin{itemize}}
\newcommand{\ei}{\end{itemize}}

\newcommand{\thet}{{{\theta}}}

\def\ga{\mathrel{\raise.3ex\hbox{$>$\kern-.75em\lower1ex\hbox{$\sim$}}}}
\def\la{\mathrel{\raise.3ex\hbox{$<$\kern-.75em\lower1ex\hbox{$\sim$}}}}

\def\l{\left}
\def\r{\right}
\def\be{\begin{equation}}
\def\ee{\end{equation}}

\def\I_M{{I_{\scriptscriptstyle M\times M}}}

\def\be{\begin{equation}}
\def\ee{\end{equation}}
\def\bea{\begin{eqnarray}}
\def\eea{\end{eqnarray}}
\newcommand{\beq}{\begin{eqnarray}}
\newcommand{\eeq}{\end{eqnarray}}

\renewcommand{\d}{\rm{d}}
\newcommand{\beqal}{\begin{eqnarray}\label}
\newcommand{\beqa}{\begin{eqnarray}}
\newcommand{\eeqa}{\end{eqnarray}}

\begin{document}
\title{\large Scalar absorption by charged rotating black holes}

\author{Luiz C. S. Leite}
\email{luizcsleite@ufpa.br}
\affiliation{Faculdade de F\'{\i}sica, Universidade 
Federal do Par\'a, 66075-110, Bel\'em, Par\'a, Brazil.}

\author{Carolina L. Benone}
\email{lben.carol@gmail.com}
\affiliation{Faculdade de F\'{\i}sica, Universidade 
Federal do Par\'a, 66075-110, Bel\'em, Par\'a, Brazil.}

\author{Lu\'is C. B. Crispino}
\email{crispino@ufpa.br}
\affiliation{Faculdade de F\'{\i}sica, Universidade 
Federal do Par\'a, 66075-110, Bel\'em, Par\'a, Brazil.}

\begin{abstract}
We compute numerically the absorption cross section of planar massless scalar waves impinging upon a Kerr-Newman black hole with different incidence angles. We investigate the influence of the black hole electric charge and angular momentum in the absorption spectrum, comparing our numerical computations with analytical results for the limits of high and low frequency.
\end{abstract}

\date{\today}

\maketitle

\section{Introduction}\label{sec:int}
The existence of black holes (BHs) has been intensively debated along many decades. Although a lot of indirect evidences that BHs are present in nature have been collected, e.g., Cygnus X-1~\cite{Cygnus}, BHs still demanded a stronger evidence of their existence. 
 The recent report of the detection of gravitational waves~\cite{GWs,Abbott:2016nmj} is remarkable, not only for the confirmation of one of the most important predictions of GR, but also because the gravitational waves detected by LIGO Scientific Collaboration are claimed to have been 
 produced by the merger of a binary BH system, providing additional evidence that BHs indeed exist in nature.

One way to investigate the nature of BHs is observing how they absorb/scatter particles and waves in 
their vicinity. In this context, many works have been done in order to determine the absorption and 
scattering cross sections of BHs, considering different BH solutions, particles and waves of 
different kind. Using numerical techniques, Sanchez investigated the absorption and scattering of 
planar massless scalar waves by a Schwarzschild BH~\cite{Sanchez:1977si,sanchez1978elastic}, taking 
a first step in this line of investigation. The absorption and scattering cross sections have also 
been studied for the case of Reissner-Nordstrom BHs~\cite{Jung:2005mr, Crispino:2008zz, 
Crispino:2009ki, Crispino:2009zza, Oliveira:2011zz, Benone:2014qaa, 
PhysRevD.90.064027, PhysRevD.92.084056, PhysRevD.93.024028, PhysRevD.95.044035}, Kerr 
BHs~\cite{Glampedakis:2001cx, 
Dolan:2008kf, Caio:2013} and regular BHs~\cite{PhysRevD.90.064001,PhysRevD.92.024012}. Recently, the 
absorption of massless scalar waves by a Schwarzschild BH surrounded by a thin shell of matter has 
also been analyzed~\cite{PhysRevD.93.024027}. However, the absorption and scattering of scalar waves 
by Kerr-Newman BHs has not been investigated yet. 

Massless scalar waves can be seen as a simpler proxy for higher-spin waves, such as the electromagnetic and gravitational waves. In addition to that, recently it was proposed that ultra-light bosons with null spin can be good candidates to dark matter \cite{PhysRevLett.85.1158,Hui:2016ltb}. Such fields, which are called fuzzy dark matter, would interact with black holes, for instance at the core of galaxies; so that it is of interest to study how these fields are scattered and absorbed by the black hole.

We can wonder if it is reasonable to assume a black hole to have a non zero electric charge. 
The electromagnetic force can lead to the discharge of a black hole, 
favoring (disfavoring)  charges of opposite (the same) sign of the 
black hole charge to be absorbed.
Another phenomenon
to take into consideration is the emission of particles by black holes~\cite{hawking1974black}. 
Charged black holes tend to emit more particles with the same charge sign 
of the black hole
than with the opposite sign, also leading to a neutralization of the black  hole \cite{Gibbons:1975kk}.
However, in the context of minicharged dark matter \cite{Davidson:2000hf} it is possible to have black holes with a considerable amount of charge and even extremally charged black holes \cite{Cardoso:2016olt}.

While investigating the scattering of fields by chargeless rotating BHs, one finds that for small enough frequencies the scattered wave can be amplified, decreasing the energy and angular momentum of the black hole. Such phenomenon is known as superradiance and it was first investigated by Misner \cite{Misner:1972kx}. Moreover, Hawking radiation brought an extra motivation to the study of the absorption cross section of BHs, since the emission rate of a BH is proportional to its absorption cross section.

In this work we consider the case of a massless scalar plane wave impinging upon a rotating and charged BH, computing the absorption cross section numerically. In the low-frequency limit we consider the general result obtained by Higuchi \cite{Higuchi:2001si} and for the high-frequency limit we consider the eikonal approach, using the geodesic equation to obtain the high-frequency absorption cross section. This paper is organized as follows: In Sec.~\ref{sec:scalarfield}, we present the Kerr-Newman metric written in Boyer-Lindquist coordinates and make a brief review of the treatment given to the massless scalar field in the Kerr-Newman background. In Sec.~\ref{sec:absorption}, we present expressions for the absorption cross section, which are obtained via partial-wave approach. In Sec.~\ref{sec:results}, we exhibit our numerical results for different choices of the plane wave incidence angle, the rotation parameter and the electric charge of the BH. We finish this paper presenting some remarks in Sec.~\ref{sec:remarks}. We adopt natural units~($c=G=\hslash=1$).

\section{Scalar field in the Kerr-Newman spacetime}\label{sec:scalarfield}
In standard Boyer-Lindquist coordinates~$\left\{t,\,r,\,\theta,\,\phi\right\}$, the Kerr-Newman BH is described by the  following line element
\bea
&\d s^2&= \l(1-\frac{2Mr-Q^2}{\rho^2}\r)\d t^2-\frac{\rho^2}{\Delta}\d r^2-\rho^2\d\theta^2 \nonumber\\
&+&\frac{4Mar\sin^2\theta-2aQ^2\sin^2\theta}{\rho^2}\d t \d\phi-\frac{\xi\sin^2\theta}{\rho^2}\d\phi^2,\label{eq:linelement}
\eea
where we have defined~$\rho^2\equiv r^2+a^2\cos^2\theta$, $\Delta\equiv r^2-2Mr+a^2+Q^2$, and $\xi\equiv (r^2+a^2)^2-\Delta a^2\sin^2\theta$. The set of parameters~($M,\,Q,\,a$) are interpreted, respectively, as the mass, the electric charge, and angular momentum per unit mass of the BH. The Kerr-Newman metric represents a BH spacetime such that: $a^2+Q^2\leq M^2$. Here we shall restrict our attention to the case~$a^2+Q^2 < M^2$, which presents two distinct horizons, namely, an internal~(Cauchy) horizon located at~$r_{-}=M-\sqrt{M^2-a^2-Q^2}$ and an external~(event) horizon at~$r_{+}=M+\sqrt{M^2-a^2-Q^2}$.

A massless scalar field $\Psi(x^\mu)$ is governed by the Klein-Gordon equation, which can be written in its covariant form as
\begin{equation}
\frac{1}{\sqrt{-g}}{\partial_\mu\l(\sqrt{-g}	g^{\mu\nu}\partial_\nu\Psi\r)}=0,\label{eq:kgeq}
\end{equation}
where $g_{\mu\nu}$ are the covariant components of the Kerr-Newman metric, which can be obtained directly from Eq.~\eqref{eq:linelement}.	$g^{\mu\nu}$ are the contravariant components of the metric and $g$ is the metric determinant. 

We can decompose the scalar field in wavelike solutions of Eq.~\eqref{eq:kgeq}, as follows:
\be
\Psi=\sum_{l=0}^{+\infty}\sum_{m=-l}^{+l}\frac{U_{\omega lm}(r)}{\sqrt{r^2+a^2}}S_{\omega lm}(\theta)e^{im\phi-i\omega t}.\label{eq:fieldecomposition}
\ee

The functions~$S_{\omega lm}$, appearing in Eq.~\eqref{eq:fieldecomposition}, are the standard oblate spheroidal harmonics \cite{abram}, which satisfy the following equation
\bea
&\l(\frac{\rm{d}^2}{\rm{d}\thet^2}+\cot\thet\frac{\rm{d}}{\rm{d}\theta}\r)S_{\omega lm}\nonumber\\
&+\l(\lambda_{lm}+a^2\omega^2\cos^2\theta-\frac{m^2}{\sin^2\theta}\r)S_{\omega lm}=0,\label{eq:spheroidaleq}
\eea
where~$\lambda_{lm}$ are the eigenvalues of the spheroidal harmonics. These angular functions are normalized as follows:
\be
\int \d\theta\,\sin\theta\,\l|S_{\omega lm}(\theta)\r|^2=\frac{1}{2\pi}.
\ee

Using the definition of the tortoise coordinate~$r_{\star}$ in the Kerr-Newman spacetime, namely,
\be
r_{\star}\equiv\int \d r\,\l(\frac{r^2+a^2}{\Delta}\r),\label{eq:tortoisecoord}
\ee
the differential equation obeyed by the radial function~$U_{\omega lm}$ can be rewritten as
\be
\l(\frac{\d^2}{\d r_\star^2}+V_{\omega lm}\r)U_{\omega lm}(r_\star)=0,\label{eq:radialeq}
\ee
with~$V_{\omega lm}$ given by
\bea
V_{\omega lm}(r)&=& \l(\omega - m\frac{a}{r^2+a^2}\r)^2\nonumber\\
&+&\l[2Mr-2r^2-\Delta+\frac{3r^2}{r^2+a^2}\Delta\r]\frac{\Delta}{\l(r^2+a^2\r)^3} 
\nonumber\\
&-&\l(a^2\omega^2+\lambda_{lm}-2ma\omega\r)\frac{\Delta}{\l(r^2+a^2\r)^2}.
\label{eq:effpot}
\eea

\begin{figure*}%
\includegraphics[width=\columnwidth]{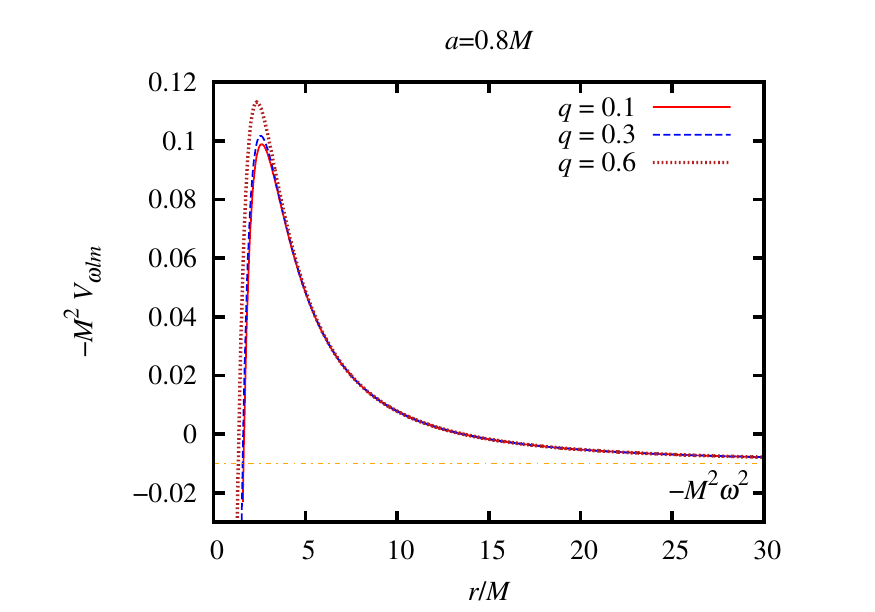}\includegraphics[width=\columnwidth]{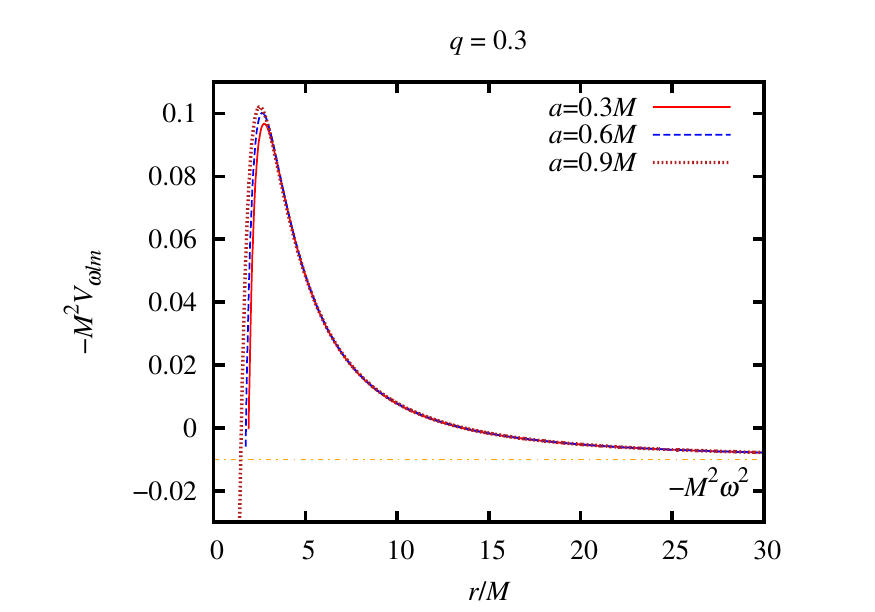}
\caption{LEFT: We plot~$-M^2V_{\omega l m}$ as function of $r$ for a fixed value of the BH angular momentum,~$a=0.8M$, and different values of BH charge-to-mass ratio, $q\equiv Q/M$. RIGHT: The function~$-M^2V_{\omega l m}$ is plotted for~$q=0.3$ and different values of~$a$. For both cases we choose $M\omega=0.1$ and $l=m=1$.}%
\label{fig:eff_pot}%
\end{figure*}

In FIG.~\ref{fig:eff_pot} we plot the function~$-M^2 V_{\omega lm}$ for different values of the BH charge-to-mass ratio, $q\equiv Q/M$, (right panel) and rotation parameter (left panel), fixing $\omega=0.1$ and $l=m=1$, in both cases. We note that the larger the BH rotation and charge are, the larger is~$-M^2 V_{\omega lm}$. Moreover, $-M^2 V_{\omega lm}$ presents a maximum and then goes to $-M^2\omega^2$ at infinity.

The radial equation~\eqref{eq:radialeq} has a set of independent solutions usually labelled as \textit{in, up, out} and \textit{down}. The \textit{in} modes are the appropriate ones to study absorption and scattering processes, since they denote purely incoming waves from the past null infinity. These modes have the following asymptotic behavior 
\be
U_{\omega lm}(r_\star)\sim\l\{
\begin{array}{c l}
	{\mathcal{I}_{\omega lm}}U_I+{{\cal R}_{\omega lm}} U_I^* & (r_\star/M\rightarrow +\infty),\\
	{\mathcal{T}_{\omega lm}} U_T & (r_\star/M\rightarrow -\infty),
\end{array}\r.
\label{inmodes}
\ee  
in which
\be
U_T = e^{-i \l({\omega-m\Omega_{\rm H}}\r)r_\star}\sum_{j=0}^N g_j (r-r_+)^j,
\ee
\be
U_I = e^{-i \omega r_\star}\sum_{j=0}^N \frac{h_j}{r^j},
\ee
and $\Omega_{\rm H}\equiv\frac{a}{r_+^2+a^2} $ is the event horizon angular velocity. 
The symbol $^*$ denotes complex conjugation.
$g_j$ and $h_j$ are constants, which can be found by assuming that Eq. (\ref{inmodes}) obeys Eq. (\ref{eq:radialeq}) in the asymptotic limits. 
The coefficients~$\mathcal{R}_{\omega lm}$ and $\mathcal{T}_{\omega lm}$ are related to the reflection and transmission coefficients, respectively, and obey the equation
\be
\l|\frac{\mathcal{R}_{\omega lm}}{\mathcal{I}_{\omega lm}}\r|^2=1-\frac{\omega-m\Omega_{\rm H}}{\omega}\l|\frac{\mathcal{T}_{\omega lm}}{\mathcal{I}_{\omega lm}}\r|^2.
\ee 	
From this relation one can see that for $0<\omega<m\Omega_{\rm H}$, $\l|{\mathcal{R}_{\omega lm}}\r|^2>\l|{\mathcal{I}_{\omega lm}}\r|^2$. This enhancement in the amplitude of the reflected wave is known as superradiance~\cite{brito2015superradiance}.

\section{Absorption cross section} \label{sec:absorption}
The absorption cross section can be well described by analytical approximate results in the low- and high-frequency regimes. In the low-frequency regime, it has been shown that the absorption cross section of massless scalar waves for stationary BHs, such as a Kerr-Newman BH, is given by the area of the event horizon \cite{Higuchi:2001si}. In the high-frequency limit, the absorption cross section tends to the capture cross section of null geodesics. These limiting results are important to check the accuracy of the numerical results, which are exhibited in Sec.~\ref{sec:results}. We will show that our numerical results, both in the low- and high-frequency regimes, are in agreement with the analytical limits.

We can obtain an expression for the absorption cross section through the partial-wave approach, so that the total absorption cross section of massless scalar waves impinging upon a Kerr-Newman BH with an incidence angle~$\gamma$, is
\be
\sigma=\sum_{l=0}^\infty\sum_{m=-l}^l\frac{4\pi^2}{\omega^2}\l|{S_{\omega lm}}(\gamma)\r|^2\l(1-\l|\frac{\mathcal{R}_{\omega lm}}{\mathcal{I}_{\omega lm}}\r|^2\r).\label{eq:abs_cs}
\ee 

According to the value assumed by the azimuthal number~$m$, we can distinguish the modes between corotating~($m>0$) and counterrotating~($m \leq 0$) with the 
BH.\footnote{Note that the $m=0$ mode has been included among the counterrotating modes.}
Hence, the total absorption cross section can be seen as a sum of the corotating
and the counterrotating contributions
allowing to split the absorption cross section into 
non-superradiant modes~($\sigma^{counter}$) 
and superradiant modes~($\sigma^{co}$)
superradiant modes~($\sigma^{co}$) 
and non-superradiant modes~($\sigma^{counter}$),
respectively~\cite{Luiz:2016}.

\subsection{Capture cross section}\label{sec:geodesic_capture}
In the high-frequency regime the absorption cross section
approaches the capture cross section of null geodesics, which is given by
\be
\sigma_{\text{geo}}=\frac{1}{2}\int_{-\pi}^{\pi}b_c^2(\chi,\gamma)d\chi,
\ee
where $\chi$ is an angle defined in the plane of the impinging wave. When $b_c$ is constant we recover $\sigma_{\text{geo}}=\pi b_c^2$, which is the high-frequency absorption cross section for static geometries.
In Ref.~\cite{Caio:2013} an approach for 
the computation of the capture cross section 
was presented, considering null geodesics impinging upon 
a Kerr BH. We can adopt the same treatment for Kerr-Newman BHs, 
obtaining basically the same formulae 
presented in Subsec. III.B.1 of Ref.~\cite{Caio:2013}.

For an impact vector with components

\be
\vec{b}=\l(b\cos\gamma \cos\chi,b\sin\chi,-b\sin\gamma\sin\chi\r),
\ee
we find,
for the radial part of the geodesic equation,
\be
R(r)=[(r^2+a^2)-a \mathcal{L}_z]^2 - \Delta[(\mathcal{L}_z-a)^2+\mathcal{K}], 
\ee
with
\be
\mathcal{L}_z=L_z E^{-1}=b \sin{\chi}\sin{\gamma}, 
\ee
and 
\be
\mathcal{K}=KE^{-2}=b^2(\cos^2\chi+\sin^2\chi\cos^2\gamma)-a^2\cos^2\gamma,
\ee
where $K$ is the Carter's constant.
In order to find the critical radius and the critical impact parameter we have to solve $R(r_c)=0$ and $R'(r_c)=0$.

\begin{figure*}
\includegraphics[width=\columnwidth]{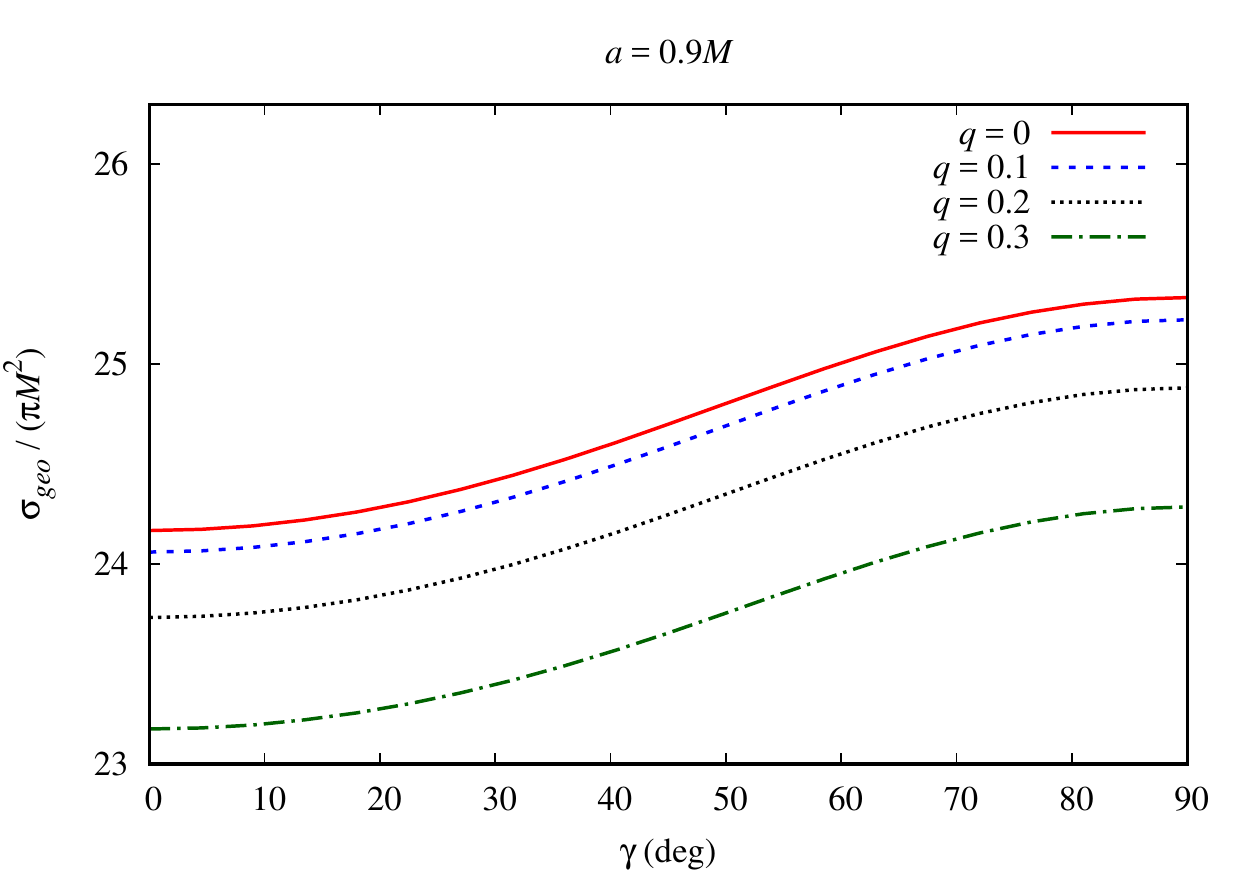}\includegraphics[width=\columnwidth]{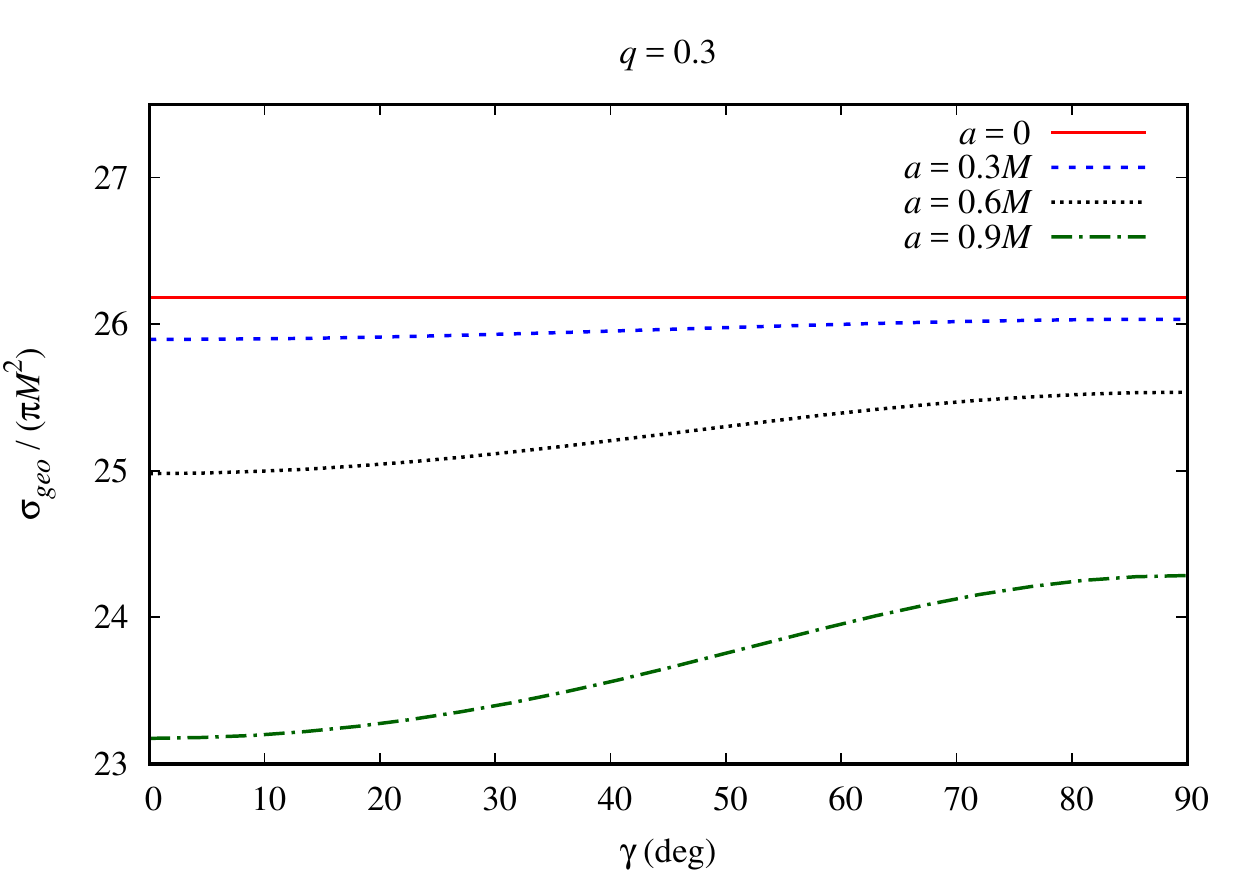}
\caption{
LEFT: The capture cross section for a fixed value of the rotation parameter~($a=0.9M$) and different 
values of the charge-to-mass ratio~($q=0$, $0.1$, $0.2$, and $0.3$). RIGHT: The capture cross 
section for a fixed value of the BH charge~($q=0.3$) and different values of the BH rotation~($a/M=0$, 
$0.3$, $0.6$, and $0.9$).}
\label{fig:geodesic_capture}
\end{figure*}

The capture cross section is exhibited in FIG.~\ref{fig:geodesic_capture} as a function of the null 
geodesic angle of incidence. We consider different values for the BH rotation parameter~(right 
panel) and electric charge~(left panel), showing that the capture cross section diminishes as we 
increase the values of the Kerr-Newman BH spin and charge-to-mass ratio. For the static case~($a=0$) the capture cross section is independent of the incidence angle, but for $a>0$ the capture cross section increases as we increase $\gamma$, reaching its maximum value at the equatorial plane~($\gamma=90$~$deg$).

\section{Results}\label{sec:results}
Along this section, we exhibit an assortment of our numerical results for the massless scalar absorption cross section of Kerr-Newman BHs. In order to obtain our results, we numerically solve Eq.~\eqref{eq:radialeq} submitted to the boundary conditions given in Eq.~\eqref{inmodes}. Knowing the radial equation solutions we are able to find the reflection and the transmission coefficients by comparing the numerical solutions with the analytical ones given in Eq.~\eqref{inmodes} at $r=r_{\infty}$, where $r_{\infty}$ represents the numerical infinity.

\begin{figure*}
\includegraphics[width=\columnwidth]{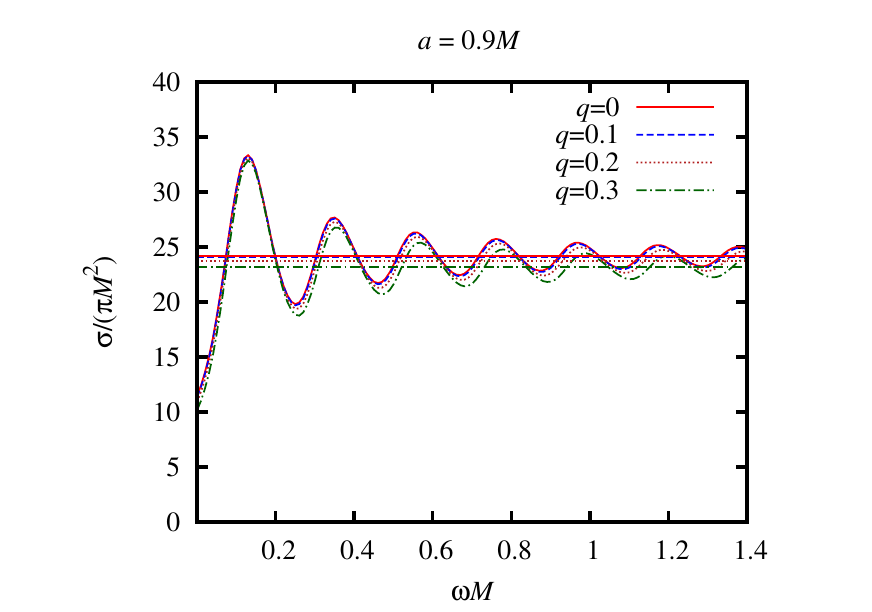}\includegraphics[width=\columnwidth]{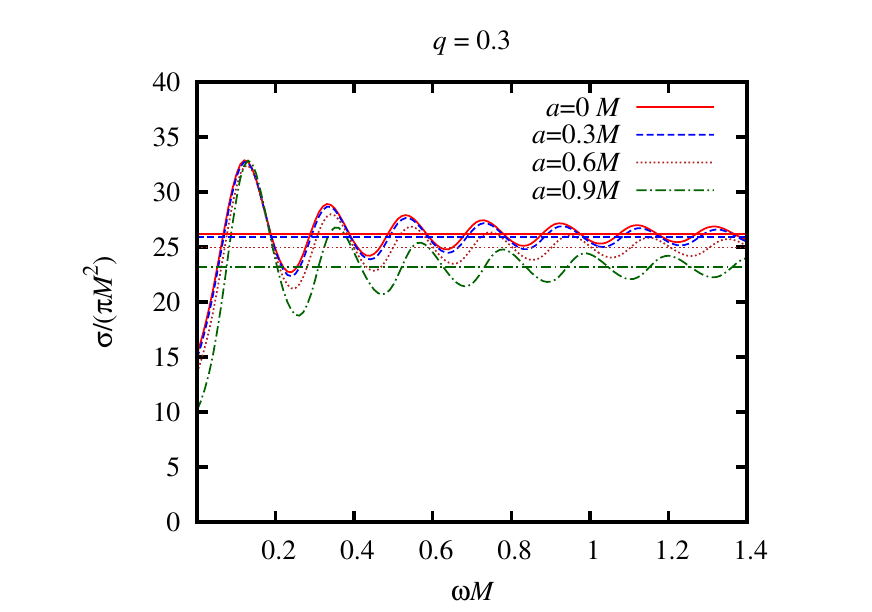}
\caption{LEFT: The total absorption cross section for a fixed BH rotation,~$a=0.9M$, with $q=0$, $0.1$, $0.2$, and $0.3$.  RIGHT: The total absorption cross section for a BH charge $q=0.3$ and different rotation parameters $a/M=0$, $0.3$, $0.6$, and $0.9$. Both left and right panels were obtained for on-axis incidences~($\gamma=0$), and the horizontal lines are the capture cross sections of null geodesics in each case.}
\label{abs_on_axis}
\end{figure*}

In FIG.~\ref{abs_on_axis} we show the total absorption cross section in the on-axis 
case~$(\gamma=0)$ for a fixed rotation parameter~(left panel) and for a fixed BH charge~(right 
panel). In the on-axis case we note that the absorption cross section presents a similar behavior to 
the static BH case, i.e., the total absorption cross section starts from a value given by its event 
horizon area and then oscillates with a decreasing amplitude around its high frequency-limit, which 
is represented by horizontal lines, approaching this value for larger values of $\omega$. 
Furthermore, we can see that the larger is the BH charge and the faster is its spin, the smaller is 
the total absorption cross section. This behavior is also observed for the capture cross section of null geodesics~(see FIG.~\ref{fig:geodesic_capture}).

\begin{figure*}
\includegraphics[width=\columnwidth]{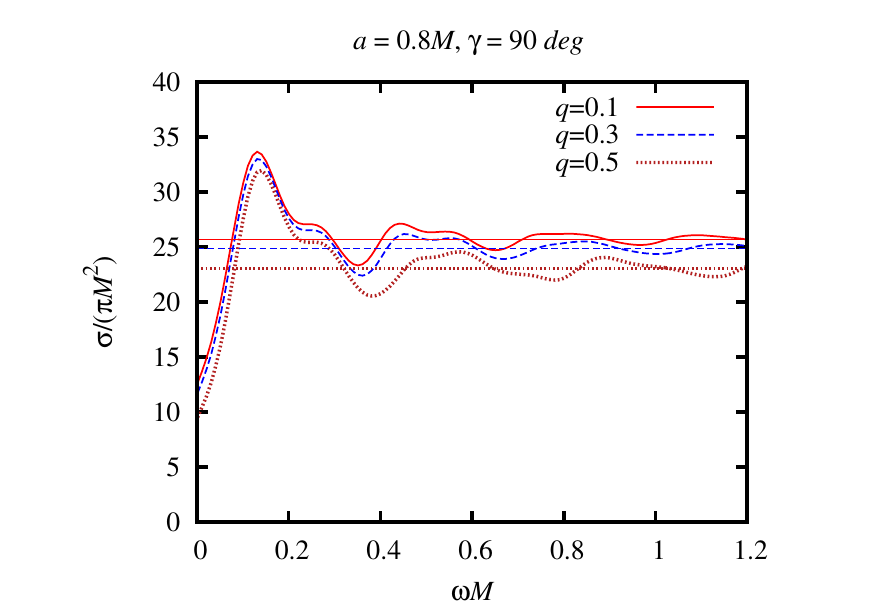}\includegraphics[width=\columnwidth]{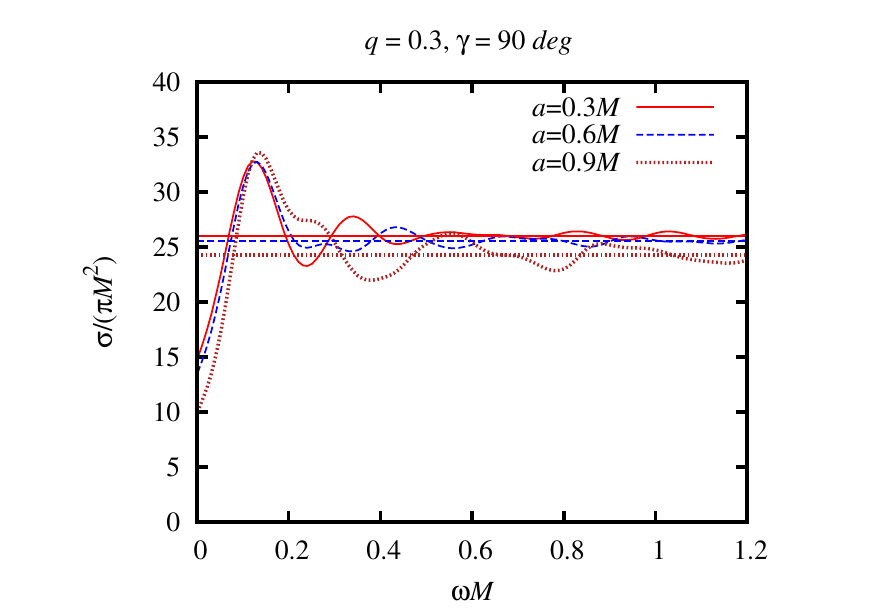}
\caption{LEFT: The total absorption cross section for incidences along the equatorial plane for~$a=0.8M$, and~$q=0.1$,~$0.3$,~and~$0.5$. RIGHT: The total absorption cross section for~$\gamma=90$  \textit{deg}, with~$q=0.3$, and different BH rotations~$a/M=0.3$, $0.6$, and $0.9$. The horizontal lines are the values of the capture cross section for null geodesics in each case.}
\label{abs_off_axis}
\end{figure*}

For off-axis incident waves the oscillating pattern of the absorption cross section is less regular than that of the on-axis case. In order to illustrate this, in FIG.~\ref{abs_off_axis} we consider incidences along the equatorial plane~$\gamma=90$ \textit{deg}. In the left panel of FIG.~\ref{abs_off_axis}, we choose Kerr-Newmann BHs with a fixed rotation parameter, $a=0.8M$, and with different charge-to-mass 
ratios~$q=0.1,\,0.3$\,and\,$0.5$. We also consider rotating and charged BHs with the same electric charge~$q=0.3$~(right panel) and different rotation parameters~$a/M=0.3,\,0.6$\,and\,$0.9$. 
As a general behavior, we can observe, in both left and right panels of FIG.~\ref{abs_off_axis}, a less regular pattern of the absorption cross section, contrasting with the case of on-axis  incidence. 
As we increase the charge-to-mass ratio, the absorption cross section decreases, but maintains its essential shape. On the other hand, as we increase the rotation, the absorption cross section becomes less regular.

\begin{figure*}
\includegraphics[width=\columnwidth]{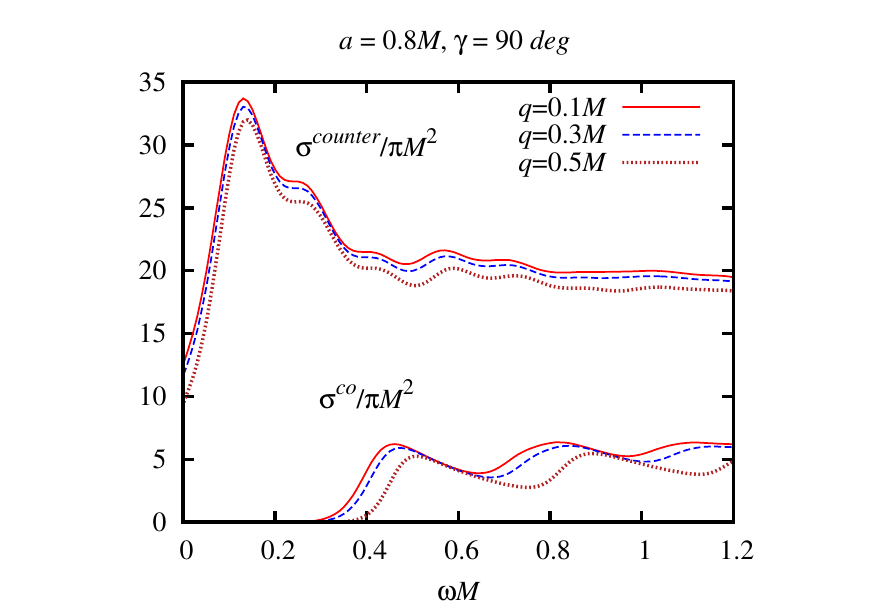}\includegraphics[width=\columnwidth]{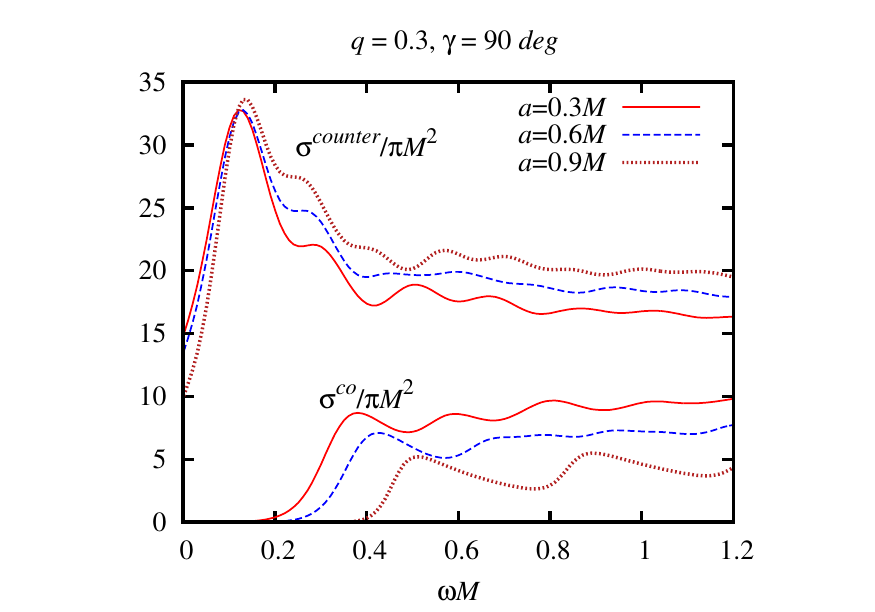}
\caption{LEFT: The corotating~($\sigma^{co}$) and counterrotating~($\sigma^{counter}$) contributions to the absorption cross section for~$a=0.8M$, and~$q=0.1$,~$0.3$, and~$0.5$. RIGHT: The corotating and counterrotating contributions to the absorption cross section for~$q=0.3M$, and~$a/M=0.3$,~$0.6$, and~$0.9$.}
\label{abs_co_cou}
\end{figure*}

In FIG.~\ref{abs_co_cou} we show results for the corotating and counterrotating contributions to the total absorption cross section. By examining the corotating and counterrotating modes separately, it is possible to identify a more regular behavior. Also, we see that the contributions of the counterrotating modes to the total absorption cross section are larger than the corotating ones. 
The corotating modes are more absorbed as either the BH charge or rotation parameter decreases. 
The counterrotating modes are more absorbed as the BH charge decreases and the behavior for different rotations depends on the frequency regime. In the low-frequency limit the absorption cross section for counterrotating waves decreases as we increase the rotation parameter. In this limit the absorption cross section is dominated by the waves with $m=0$ and goes to the area of the black hole as $\omega\rightarrow 0$. In the high-frequency regime, $\sigma^{counter}$ increases as $a$ increases.

\begin{figure*}
\includegraphics[width=\columnwidth]{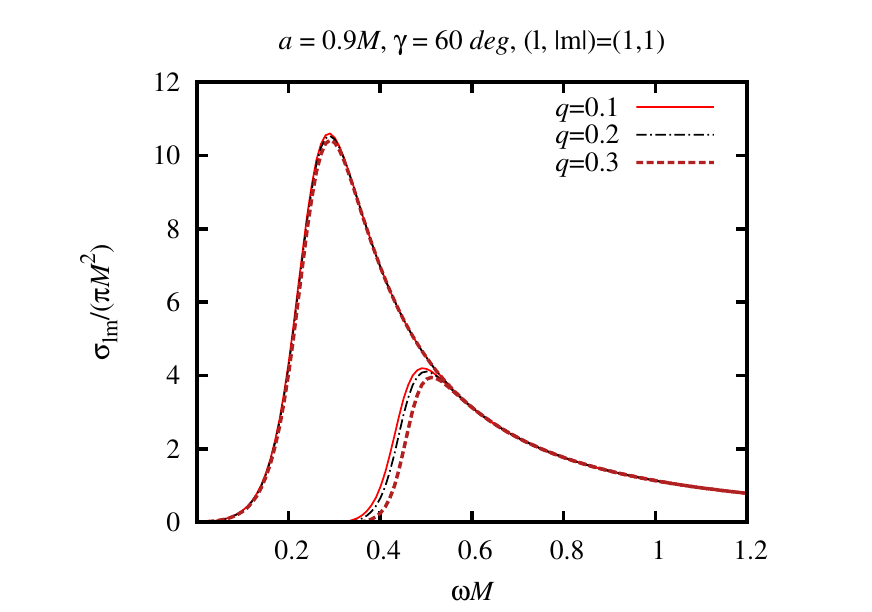}\includegraphics[width=\columnwidth]{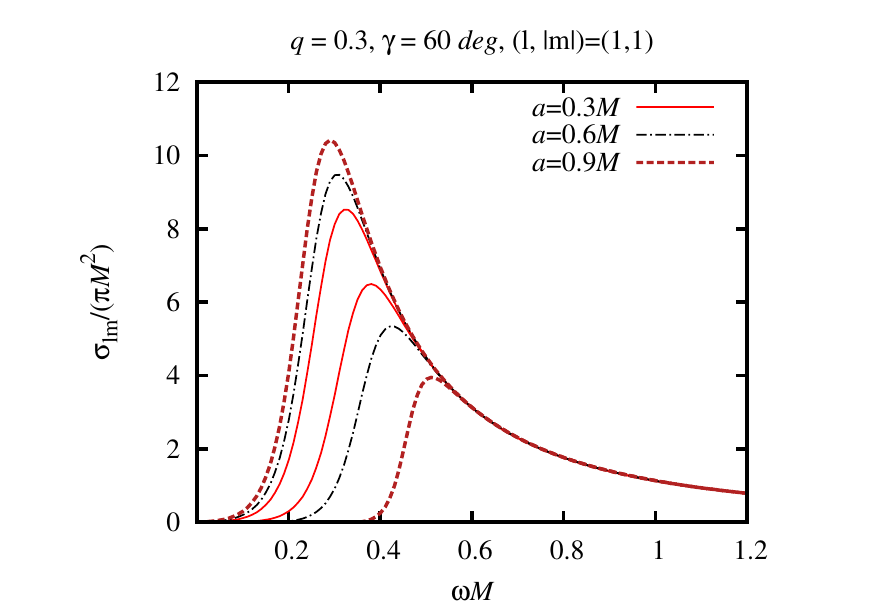}
\caption{LEFT: Counterrotating~($m=-1$) and corotating~($m=1$) partial absorption cross sections for~$\gamma=60~\rm{deg}$, $a=0.9M$, and $q=0.1$,~$0.2$, and $0.3$. RIGHT: Counterrotating and corotating partial absorption cross sections for a fixed BH charge~($q=0.3$), and different rotation parameters~$a/M=0.3$,~$0.6$, and~$0.9$. We consider $\gamma=60~\rm{deg}$. In both right and left panels, the curves associated to the higher peaks are related to the counterrotating modes while the low peaks are associated to corotating ones. }
\label{partial_lm11}
\end{figure*}

In FIG.~\ref{partial_lm11} we present the partial absorption cross sections for the dipole mode~($l=1$). As the BH charge increases, both the corotating~($m=1$) and counterrotating~($m=-1$) partial cross sections decrease (left panel). On the other hand, as the rotation parameter increases  the associated values of the corotating partial absorption cross section decrease, while the counterrotating ones increase (right panel). 

\begin{figure*}
\includegraphics[width=\columnwidth]{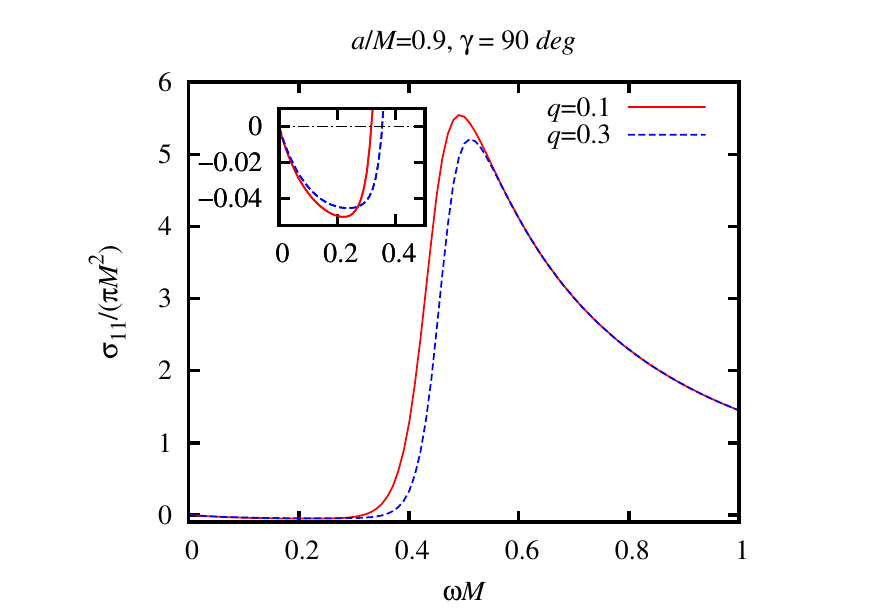}\includegraphics[width=\columnwidth]{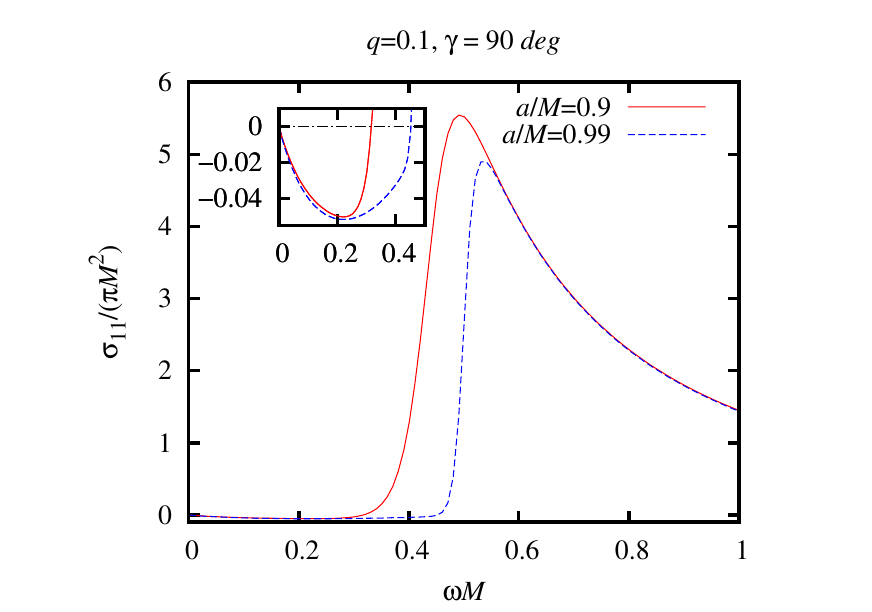}
\caption{LEFT: The partial absorption cross section for the mode~$l=m=1$ with $a=0.9M$, and $q=0.1$ and $0.3$. RIGHT: The partial absorption cross section for the mode~$l=m=1$, with $q=0.1$, and $a/M=0.9$ and $0.99$. In both left and right panels, the insets help to see more clearly the superradiance, which is very small for scalar waves.}
\label{zoom_lm11_gam90}
\end{figure*}

In the Kerr-Newman spacetime, due to superradiance, reflected waves can be amplified, what results in a negative partial absorption cross section. We show the partial absorption cross sections, in FIG.~\ref{zoom_lm11_gam90}, for the mode~$l=m=1$, considering incidence along the equatorial plane, case for which superradiance is more evident. For a fixed rotation parameter~(left panel), the larger the BH charge is, the smaller is superradiance. On the other hand, as the rotation parameter increases, superradiance increases (right panel).

\section{Final remarks}\label{sec:remarks}
We have obtained numerically the absorption spectrum of planar massless scalar waves for a Kerr-Newman BH, considering different values for the BH rotation parameter and electric charge. 
We have confirmed with our numerical results that in the low-frequency regime the absorption cross section 
tends to the area of the BH horizon, while in the high-frequency regime it approaches 
the capture cross section of null geodesics.

We have shown that the absorption cross section presents a regular oscillatory behavior around its high-frequency limit when the waves impinge along the rotation axis~($\gamma=0$). This oscillatory behavior comes from the contributions of waves with different angular momentum, $l$. As either the charge or the rotation parameter increases, the absorption cross section decreases, what is consistent with the fact that when we increase these parameters both the area and critical impact parameter of the black hole decrease.

For off-axis incidences~($\gamma\neq 0$) we have observed 
that the absorption cross section behaves with less regular oscillations around the geometric capture cross section, what is a consequence of the different contributions given by the co- and counterrotating modes. We have also noted that, for off-axis incidences, the oscillatory pattern becomes more irregular for larger values of the BH rotation parameter.

When we consider the corotating and counterrotating contributions to the absorption cross section separately, 
we identify a more regular absorption profile, 
and we observe that the counterrotating modes are more absorbed than the corotating ones. 
Moreover, we have obtained that the larger the BH charge and rotation parameter are, the less absorbed are the corotating waves. For the counterrotating case we have obtained that the waves are more absorbed as the BH charge decreases, and as we increase the rotation parameter we have different behaviors for different frequency regimes. This happens because in the low-frequency limit the field is dominated by the modes with $m=0$, tending to the area of the BH horizon, which decreases as we increase the BH rotation. 
As we reach higher values of the frequency 
the absorption cross section increases for bigger values of the BH rotation parameter.

For the case of superradiant scattering 
we found that superradiance is larger 
the faster the BH spins and the smaller is the BH charge. 
As for the Kerr case, superradiance yields a 
negative partial absorption cross section, 
although the total absorption cross section remains 
positive.

\begin{acknowledgments}

The authors would like to thank Sam R. Dolan for discussions and 
acknowledge 
Conselho Nacional de Desenvolvimento Cient\'ifico e Tecnol\'ogico (CNPq)
 and Coordena\c{c}\~ao de Aperfei\c{c}oamento de Pessoal de N\'ivel Superior (CAPES) 
 for partial financial support.
 L. L. thanks the University of Sheffield for kind hospitality.
 
\end{acknowledgments}

\bibliography{refs}

\end{document}